\documentclass[twocolumn]{article}
\usepackage{amsmath,amsthm,amssymb}
\usepackage{graphicx,color}
\usepackage{wrapfig}
\makeatletter
\def\@normalize{@setsize\normalsize{12pt}\xpt\@xpt
\abovedisplayskip 10pt pluse2pt minus5pt\belowdisplayskip
\abovedisplayskip \abovedisplayshortskip \z@ plus3pt\belowdisplayshortskip
6pt plus3pt minus3pt\let\@listi\@listI}
\def\subsize{\@setsize\subsize{12pt}\xipt\@xipt}
\makeatother

\begin{document}
\date{}
\title{\large\bf Universality in Globally Coupled Maps and 
Flows}
\author{Tokuzo Shimada$^1$, Takanobu Moriya$^1$, Hayato Fujigaki$^2$\\ 
$^1$Department of physics, School of Science and Technology, Meiji University \\
1-1-1 Higashimita, Tama, Kawasaki, Kanagawa 214-8571\\
$^2$Hitachi Software Engineering Co.,Ltd. \\
4-12-7 Higashi Shinagawa 
Shinagawa, Tokyo 140-0002}
\maketitle
\thispagestyle{empty}
\subsection*{\centering Abstract}
\vspace*{-3mm}
We show that universality in chaotic elements 
can be lifted to that in complex systems. We construct a globally coupled Flow lattice (GCFL), 
an analog of GCML of Maps. We find that Duffing GCFL shows the same behavior with GCML; 
population ratio between synchronizing clusters acts as a bifurcation parameter. 
Lorenz GCFL exhibits interesting two quasi-clusters in an opposite phase motion. 
Each of them looks like Will o' the wisp; they dance around in opposite phase.
\\
{\it Keyword : synchronization, universality, globally coupled maps and flows, Lorenz model, Duffing oscillator}
\section{Introduction}
In studying chaotic systems we may consider two facts as guiding principles.
One is `Universality in Chaos'\cite{cvitanovic}. 
Especially we here consider the universality between a map (e.g. a logistic map, a circle map
$\cdots$) and a flow, that is a system described by an ordinary differential equation 
(a Duffing oscillator, Lorenz flow $\cdots$).    
The other is a self-organization of a certain attractor by synchronization\cite{pikovsky} 
in a complex system under conflict between randomness and coherence. 
One impressive model embodying this phenomenon is 
Globally Coupled Map Lattice (GCML)\cite{kkaneko},\cite{tskk}. In this paper we study 
Globally Coupled Flow Lattice (GCFL)\cite{fns} and show that the universality at the level of the 
constituents may be extended to the level of the whole system, that is, our GCFL has much 
in common with GCML.
\section{Universality in Chaos}
Let us first briefly recapitulate the universality in chaos 
at the level of elements\cite{cvitanovic}.
Perhaps the simplest chaotic system is an iterated logistic map 
$x_{n + 1} = f_a (x_n )$
with a quadratic map $f_a(x) = 1 - ax^2$ and a nonlinear parameter $a$ in the 
range $0 < a \le 2$.
Another simple system is an iterated circle map where 
the map is replaced by $g_{b}(x)=b \sin(\pi x)$.
Both systems share the same Feigenbaum ratio $4.699\cdots$ and subject to the same universality
class. This is a piece of the universality in chaos, which was beautifully 
explained by Feigenbaum and by Cvitanovi\'{c} using renormalization 
group\cite{cvitanovic}, \cite{feigenbaum}. 

Next let us look at a system
described by an ordinary differential equation (a flow for brevity).
The Lorenz model, for instance, is described by a three dimensional ODE
\begin{eqnarray}
\label{Lorenz_equation}
\frac{dx}{dt}=\sigma(y-x),~\frac{dy}{dt}=-xz+rx-y,~\frac{dz}{dt}=xy-bz.
\end{eqnarray}
 We here consider the range of $r$ from 230 to 200 ($\sigma=10$, $b=\frac{8}{3}$),
where the attractor bifurcates from a limit cycle to chaos. 
As for the 2-dimensional flow, we consider a Duffing oscillator,
\begin{equation}
\label{duffing-renritu}
\frac{dx}{dt}=y, ~\frac{dy}{dt}=-ky-x^3+A\cos(t).
\end{equation}
With $A=0.75$ and decreasing $k$, the final attractor again bifurcates from a limit cycle to chaos. Above three models exhibit the same bifurcation structure as well as periodic windows. 
The reason why is essential for our study\cite{cvitanovic}. 
\begin{wrapfigure}{r}{25mm}
\begin{center}
 \includegraphics[width=30mm,clip]{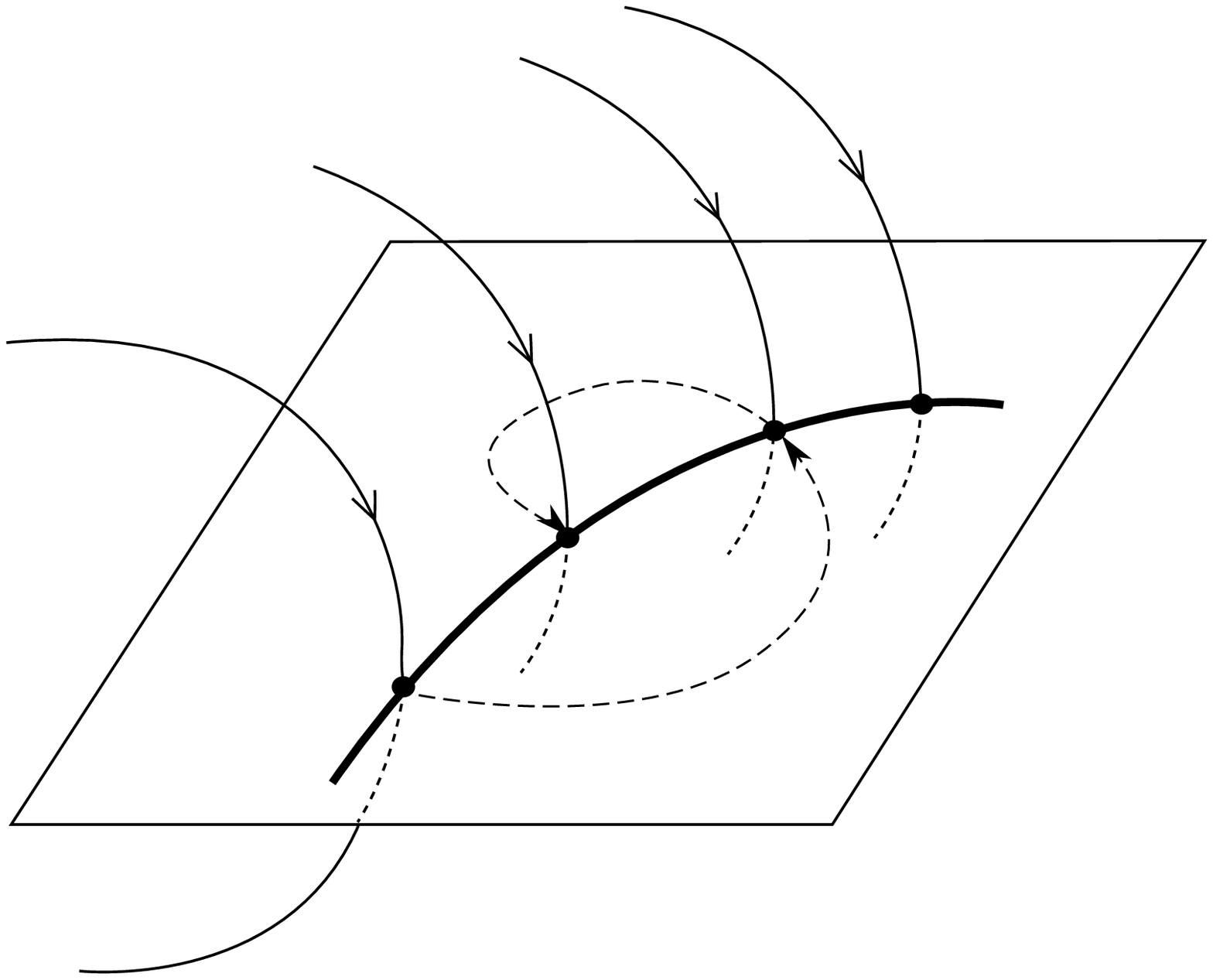}
 \caption{}
\vspace{-8mm}
\label{fig:poincare_english_version}
\end{center}
\end{wrapfigure}
The attractor of Lorenz model reduces to two dimensional due to dissipation. Then, the Poincar\'{e} section of Lorenz flow (that of Duffing oscillator) is nothing but 
one-dimensional iterated map, see Fig.~\ref{fig:poincare_english_version}. That is, if the one-dimensional map bifurcates, the corresponding flow must also bifurcates. And then, at the bifurcation limit, one-dimensional 
maps (in the same class) are all governed by a universal function. Therefore, we can amazingly understand the route from convection to turbulence from a simple logistic map.

\section{GCML}

 The GCML has $N$ maps as elements and evolves under all to all
interaction via their mean field. Here we consider its simplest form, where 
the map is a logistic map with the same $a$ and the coupling is also 
common (homogeneous GCML)\cite{kkaneko}\cite{tskk}. It evolves in an iteration of two steps. The first is a parallel mapping 
\begin{equation}
\label{eq3}
x^{\rm mid}_i \equiv f_a\left(x_{i}(n)\right)=1-a x_i^2(n),~i=1,\cdots,N
\end{equation}
where each map evolves with high nonlinearity and randomness is introduced in 
 the  system. The mean field is then calculated as 
$h(n) \equiv \sum_{j=1}^{N}x^{\rm mid}_j /N$. The second step is interaction
via the mean field 
\begin{equation}
x_{i}(n+1) = (1-\varepsilon)x^{\rm mid}_{i}+\varepsilon h(n),~i=1,\cdots,N
\end{equation}
with a coupling constant $\varepsilon$. In this step all the maps are 
pulled to the mean field and coherence is introduced in the system.
\begin{figure}[!htbp]
\begin{center}
  \includegraphics[height=40mm,clip]{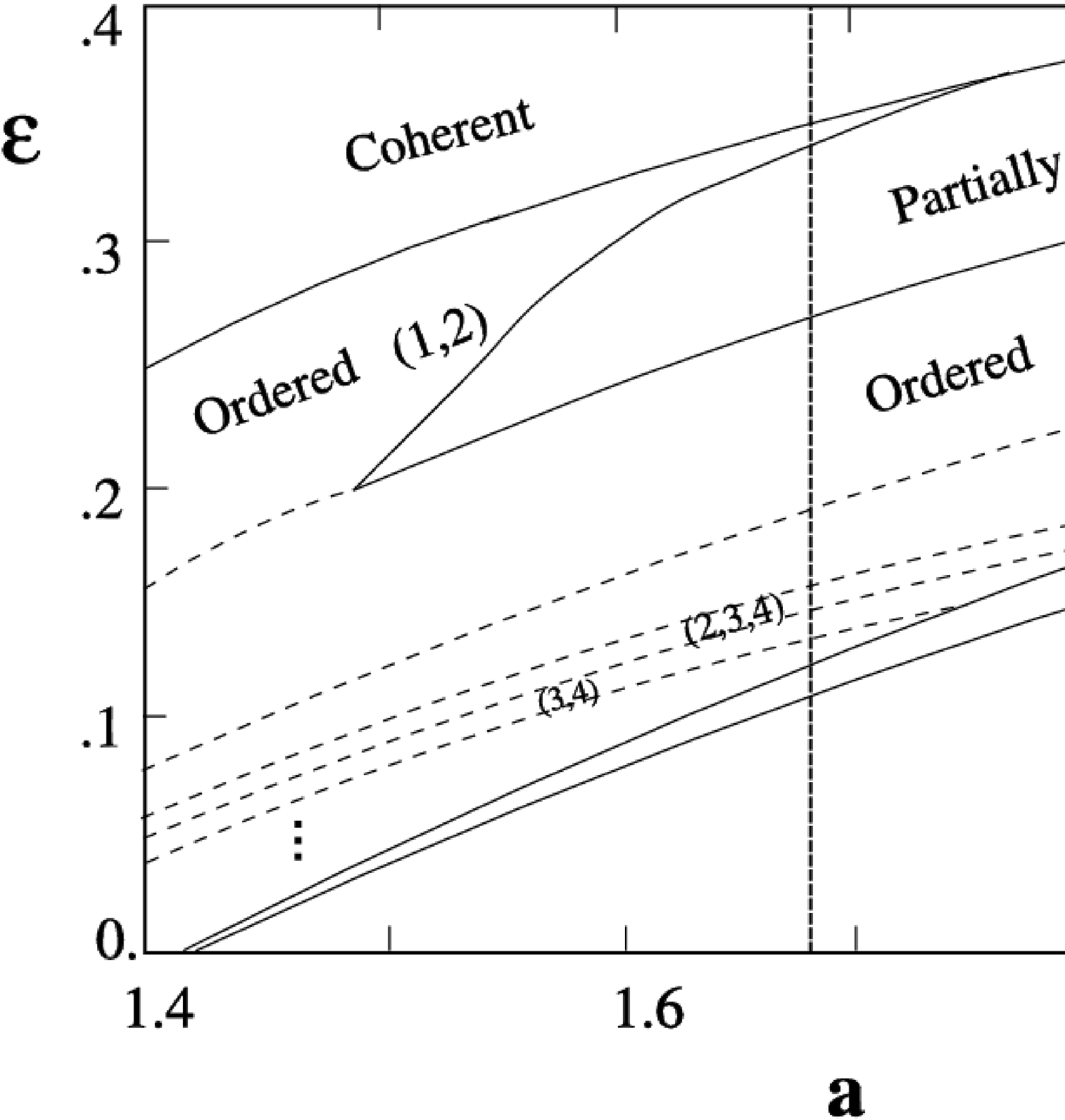}
  \caption{}
\vspace{-5mm}
\end{center}
  \label{gcml_phase_diagram}
\end{figure}
Under the conflict of randomness and coherence GCML shows various interesting phases. The phase diagram, explored first by Kaneko\cite{kkaneko} is shown in Fig.~\ref{gcml_phase_diagram}. 
We should note that the region with a very small $\varepsilon$ is called the turbulent
regime, but actually there occur drastic global periodic motions of maps,
if $\varepsilon$ takes certain values for a given $a$. 
The turbulent regime is controlled by the periodic windows
and their foliations\cite{tskk}. 

\section{Duffing GCFL}

We transport the basic structure of GCML, the iteration of independent 
evolution and subsequent interaction, to GCFL. We choose as elements $N$ Duffing oscillators that are discretized in time. 
The first step is 
\begin{equation}
\label{evolution}
\begin{aligned}
x_{(i)}^{\rm mid}&=x_{(i)}(t)+y_{(i)}(t)\varDelta t \\
y_{(i)}^{\rm mid}&=y_{(i)}(t)+\left(-ky_{(i)}(t)-\left(x_{(i)}(t)\right)^3+7.5\cos(t)\right)\varDelta t
\end{aligned}
\end{equation}
and the second is
\begin{eqnarray}
\begin{aligned}
x_{(i)}(t+\varDelta t)&=(1-\varepsilon)x_{(i)}^{\rm mid}+\frac{\varepsilon_D}{N}\sum_{i=1}^{N}x_{(i)}^{\rm mid} \\
y_{(i)}(t+\varDelta t)&=(1-\varepsilon)y_{(i)}^{\rm mid}+\frac{\varepsilon_D}{N}\sum_{i=1}^{N}y_{(i)}^{\rm mid}. 
\label{interaction}
\end{aligned}
\end{eqnarray}
 In order to investigate the phase structure of Duffing GCFL, we need
some rough estimate of the starting point in the parameter space, since otherwise 
$\varepsilon_D$, for instance, can be any value from $0$ to $1$. 
In GCML the two clustered phase in opposite phase is the most 
remarkable state formed by synchronization and realized for $\varepsilon \approx 0.19-0.27$ at $a\approx 1.68$. Let us start estimating corresponding values of Duffing parameters $k$ and $\varepsilon_D$. 
Whether corresponding cluster state is realized in Duffing GCFL
or not is of course highly nontrivial.

A natural way to estimate $k$ 
is to use the universality at the element level and match the bifurcation tree 
of the Duffing oscillator with that of the logistic map
by a scale transformation\footnote{
Universality in bifurcation is strictly a valid concept at the limit of sequential bifurcation, but in practice a simple scaling makes the two trees 
overlap each other very well.}.
We find that Duffing $k \approx 0.24$ corresponds to logistic $a\approx 1.68$.

As for $\varepsilon_D$, 
we should notice that the averaging interaction is introduced at every $\Delta t$ in GCFL. 
Hence ,if the period is $T$, the estimate may be given by
\begin{equation}
\label{Map_flow}
 (1-\varepsilon_{D})^{\frac{T}{2 \varDelta t}}=1-\varepsilon_{\rm map}.
\end{equation}
The inclusion of factor 2 in the denominator needs short explanation. 
The two clusters in GCML are oscillating in opposite phase
and at every two steps each comes back to previous value.
\begin{eqnarray*}
(+-+-+-\cdots)&\longrightarrow& + \\
(-+-+-+\cdots)&\longrightarrow& - 
\end{eqnarray*}
Therefore two steps (one step) in GCML correspond to one period (half period) 
in Duffing GCFL.
With $T =2 \pi$ and $\varDelta t=10^{-3}$, we obtain
\begin{equation}
   6.6\times10^{-5} \lesssim \varepsilon_{\rm flow} \lesssim 9.7\times 10^{-5}.
\label{epsilonestimate}
\end{equation}

We should add that we have approximated the sequence of two steps, Eq.(\ref{evolution}) and Eq.(\ref{interaction}), namely 
$\left(\text{evolution} \cdot \text{interaction}\right)^{T/\Delta t}
$
by 
$(\text{evolution})^{T/\Delta t} \cdot (\text{interaction})^{T/\Delta t}
$
in estimating $\varepsilon_D$. 

\subsection{Phase Structures}

\begin{figure}[!htbp]
 \begin{center}
  \includegraphics[height=45mm,clip]{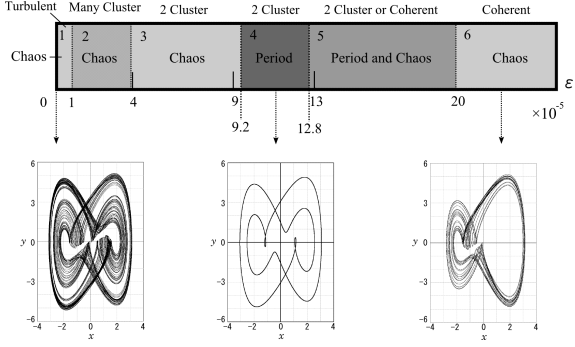}
  \caption{Upper:~Phase diagram Duffing GCFL with respect to $\varepsilon$ at $A=7.5,~k=0.24$. Lower:~attractors of he model ($N=20$) in three typical region; turbulent, two clustered state, coherent chaos with increasing $\varepsilon$.}
\label{gcduffing_phase_diagram}
 \end{center}
\end{figure}

In Fig.~\ref{gcduffing_phase_diagram} we show the phase structure  
of Duffing GCFL with respect to $\varepsilon$
at $A=7.5$ and $k=0.24$.
Let us explain the observed states in order. 
\begin{itemize}
\item [(1)]Turbulent phase. At very small $\varepsilon$, each flow evolves almost independently in a chaotic orbit.  
\item [(2)]Many clusters in chaos. Flows divide into many clusters and each cluster evolves chaotically. 
\item [(3)] Two chaotic clusters. The same as above but the number of clusters is only two.
\item [(4)] Two periodic clusters. Flows divide into two clusters and each cluster evolves periodically in approximately opposite phase to the other.
\item [(5)] Two periodic clusters or coherent chaos depending on the initial values of flows 
\footnote{
This is because Duffing oscillator has two possible attractors
mutually symmetric with respect to the origin of the $xy$ plane.
In the coherent chaos case, all flows are attracted to one of the attractors,
while in the 2 clusters case, flows divide themselves into two groups,
one for each attractor.}.   
\item [(6)] Coherent chaos. All flows at strong coupling bunch together and move in a chaotic orbit.
\end{itemize}
To our knowledge, this is the first time observation of 
the fourth phase (dark shaded region in Fig.~\ref{gcduffing_phase_diagram})---periodic clusters of flows in opposite phase motion just in accord with GCML
\cite{ieice}. 
We further study this phase below. But not only this but also 
the global phase structure is in agreement with GCML. 
This is a case that universality between elements 
(a logistic map $\sim$ a Duffing oscillator) 
also holds between systems (logistic GCML $\sim$ Duffing GCFL).
Note that the range for the two clustered phase 
is 
\begin{equation}
9.2\times10^{-5} \lesssim \varepsilon_\text{D} \lesssim 12.8\times10^{-5}
\label{Duffing-estimate}
\end{equation}
as seen in Fig.~\ref{gcduffing_phase_diagram}. 
This overlaps with (\ref{epsilonestimate}) but is shifted to the larger. Considering that our estimate is rather rough, this agreement is 
remarkable.

\subsection{Two Clustered Phase}

In GCML this phase is a consequence of reduction of nonlinearity 
due to the averaging interaction\cite{tskk,shibatakaneko,perezcerdeira}
and the fluctuation of the mean field is minimized for the stability of the 
system. 
Further interesting property of the two clustered attractor in GCML is that
it may be controlled by changing the population ratio of two
clusters\cite{kkaneko,tsstarob}. Let us call $N_{+}$ ($N_{-}$) the number of maps beyond the 
mean field at some iteration step (say $n=1000$) after random start.
Then the attractor of GCML is determined if we give not only $a$ and $\varepsilon$ 
but also $\theta\equiv N_+/N$.

\begin{figure}
 \begin{center}
  \includegraphics[width=70mm,clip]{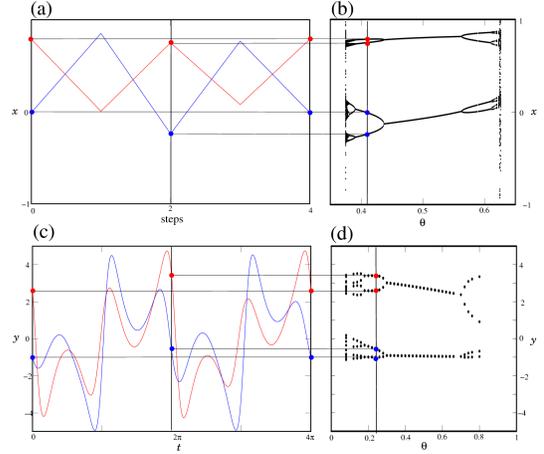}
  \caption{Bifurcation of final attractors in two clustered phase. 
  Upper: Logistic GCML with $a=1.68,~\varepsilon=0.23$ and $N=10^4$. 
  (a) Evolution of two clusters ($\theta=0.41$). (b) bifurcation of cluster 
      attractor with respect to $\theta$. Maps are plotted for even $n$ 
    ($n=1000, 1002, \cdots$).	   
  Lower: (c) Duffing GCFL with $A=7.5,~k=0.24,~\varepsilon=1.1 \times 10^{-4}$
  and $N=10^4$. ($\theta=0.22$) (d) The same with (b) but for GCFL. Poincar\'{e} shot of flows are taken at every $\Delta t=2\pi$.}
  \label{fig:forks_and_evolution}
 \end{center}
\end{figure} 

In Fig.~\ref{fig:forks_and_evolution}(a) we show 
how the GCML two clusters evolve 
when they are formed with fixed $\theta$ from random start 
and in (b) how their orbits change with $\theta$ by plotting 
all maps $x_i(n)$ at every even $n$ in order to separate the two 
clusters.  Correspondingly
we show in (c) the evolution of GCFL clusters with fixed $\theta$ 
and in (d) the bifurcation structure of the cluster
orbits by plotting the Poincar\'{e} shot of flows at every $2\pi$. 
We see clearly that both GCML and GCFL cluster attractors
share the same bifurcation structure with the variation of $\theta$.
In more detail, GCML at $even$ iteration steps and Duffing GCFL at 
every $\Delta t=2\pi$ are in one to one correspondence.

\section{Lorenz GCFL}

The Lorenz GCFL is constructed just in the same way with 
Duffing GCML (the iteration of two step process, discretized evolution in 
$\Delta t$ and subsequent interaction via the mean field).
The model parameters are now $r$ and $\varepsilon_D$ with 
$\sigma=10$ and $b=8/3$ fixed. 
We are interested in the self-formation of cluster structures
so we use large $N$, typically $N=10^4$.

By matching the Lorenz tree with the logistic tree
we estimate that $r=208$ for Lorenz flow corresponds to logistic $a=1.68$.
At $r=208$, the one turn of Lorenz chaotic attractor is roughly 
$T=0.50$. With $\varDelta t=10^{-3}$, the correspondence (\ref{Map_flow}) 
gives $\varepsilon_\text{L} \approx 8.2 \times 10^{-4}-12.0\times10^{-4}$. 
It turns out that there occur interesting quasi-periodic cluster attractors 
for $\varepsilon_\text{L}$ both at about 50 percent 
and about twice of this range, while at $\varepsilon_\text{L}$ within this range there occurs transitive behavior between the two. 
For further small $\varepsilon_L$, flows evolve almost randomly, while 
for further large $\varepsilon_L$, all flows bunch together and show
coherent chaos with the high nonlinearity of the element at $r=208$.
This is just the same way with the GCML and Duffing GCFL.
We stress that only with the help of our estimate from universality, 
we can access these interesting three regimes.

\begin{figure}[!htbp]
\begin{center}
  \includegraphics[width=80mm,clip]{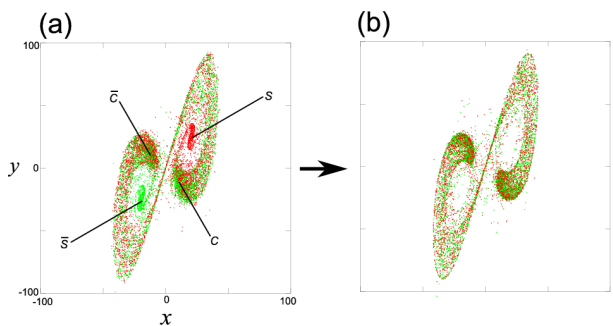}
\caption{Formation of clusters in Lorenz GCFL with
$\varepsilon=24\times10^{-4}$, $r=208$, $N=10000$, $\Delta t=10^{-3}$. 
(a)~$t=3$, (b)~$t=30$. }
\label{gclorenz_strong_formation_process}
\end{center}
\end{figure}

Let us investigate the strong coupling regime in detail\footnote{Other regimes
will be discussed elsewhere.}. Here we find interesting behavior 
of quasi-spatio clusters;  each of them, if we may say, resembles will-o'-the-wisp very much.
In Fig.~\ref{gclorenz_strong_formation_process} 
we show the final stages of their formation process (projecting into the $xy$ plane) 
from a random start at the central $\varepsilon_L$ of this regime ($\approx 24 \times 10^{-4}$). 
At around (a) ($t \approx 3$) most of flows divide themselves into two large spatial clusters 
C and $\overline{\text{C}}$ and in addition two small clusters s and $\overline{\text{s}}$ 
with high population density are formed. 
In the $xy$ projection C and $\overline{\text{C}}$ (s and $\overline{\text{s}}$) 
locate oppositely each other with respect to the origin reflecting the symmetry of Lorenz flow. 
They circulate around periodically in the first and third quadrant respectively 
(always opposite in phase). Hence, the fluctuation of the mean field in the $xy$ direction 
is suppressed. 
The clusters C and $\overline{\text{C}}$ actually exchange some part  
of their members at every time when they come close each other 
in circulation. Always the spatial clusters C and $\overline{\text{C}}$ exist 
but members are gradually mixed up. 
In this sense C and $\overline{\text{C}}$ may be called as quasi-clusters. 
(On the other hand cluster s and $\overline{\text{s}}$ do not mix.)    
Around $t=25$ (after about 100 circulations in the respective quadrant) 
the smaller cluster s (and $\overline{\text{s}}$) disappear. 
The quasi-clusters C and $\overline{\text{C}}$ remain circulating opposite in phase.

\section{Conclusion}

We have constructed GCFL, matching the nonlinearity parameter and 
the coupling with those of GCML.
The matching is only a necessary condition for the GCFL 
to inherit the properties of GCML. 
We summarize below to what extent the intriguing features of GCML
are realized in the GCFL.

Duffing GCFL preserves all of features of GCML, except some shift 
of the coupling $\varepsilon_D$ to the larger side than the prediction.
The final attractor is remarkably controlled by the population ratio $\theta$
just in the same way with GCML.
On the other hand the flows of Lorenz GCFL do move 
on a two dimensional surface but they do not form a tightly bound clusters. 
They form quasi-clusters (with mixing) in opposite phase motion 
at the strong coupling regime.


\begin{thebibliography}{9}

\bibitem{cvitanovic} P. Cvitanovi\'{c}, in {\it Universality in Chaos}, 2nd edition, Institute of Physics Publishing, Bristol and Philadelphia  
(1996).


\bibitem{pikovsky}
A. S. Pikovsky, and J. Kurths, Phys. Rev. Lett. {\bf 72}, pp. 1644-1646 (1994).

\bibitem{kkaneko} K. Kaneko, Phys. Rev. Lett. {\bf 63}, pp. 219-223 (1989).

\bibitem{tskk} T. Shimada and K. Kikuchi, Phys. Rev.  \textbf{E62}, pp. 3489-3503 (2000). 

\bibitem{fns}
H. Fujigaki, M. Nishi, and T. Shimada, Phys. Rev. {\bf E53}, pp. 3192-3197 (1996); 
H. Fujigaki and T. Shimada, Phys. Rev. {\bf E55}, pp. 2426-2433 (1997).

\bibitem{feigenbaum} M. J. Feigenbaum, Los Alamos Science {\bf 1}, pp. 4-9 (1980).

\bibitem{ieice} 
An early work is reported in 
T. Shimada, IEICE technical report, 
Nonlinear problems, 
{\bf 97}, No. 591, pp. 71-79 (1998).

\bibitem{tsstarob} T. Shimada and S. Tsukada, Proc. of the 6th Int. 
Symp. on Artificial Life and Robotics (AROB 6) {\bf 1} pp. 242-245 (2001).

\bibitem{shibatakaneko} T. Shibata and K. Kaneko, Physica D {\bf 124} pp. 177-200 (1998).

\bibitem{perezcerdeira}
G. Perez and H. A. Cerdeira, Phys. Rev. {\bf A46}, pp. 7492-7497 (1992).
\end{thebibliography}
\end{document}